\documentclass[prd,letterpaper,onecolumn,amsmath,amsfonts,amssymb,nofootinbib]{revtex4}
\pdfoutput=1
\usepackage {amsmath}
\usepackage[dvips]{graphicx}
\usepackage{feynmp}
\usepackage{amssymb}
\newcommand{\be}{\begin{equation}}
\newcommand{\ee}{\end{equation}}
\newcommand{\bear}{\begin{eqnarray}}
\newcommand{\eear}{\end{eqnarray}}

\newcommand{\lapproxeq}{\lower .7ex\hbox{$\;\stackrel{\textstyle
<}{\sim}\;$}}
\newcommand{\gapproxeq}{\lower .7ex\hbox{$\;\stackrel{\textstyle
>}{\sim}\;$}}
\newcommand{\stackdown}[2]{\lower 1.4ex\hbox{$\;\stackrel{\textstyle{#1}}
{\scriptstyle{#2}}\;$}}
\newcommand{\beq}{\begin{equation}}
\newcommand{\eeq}{\end{equation}}

\newcommand{\ba}{\begin{eqnarray}}
\newcommand{\ea}{\end{eqnarray}}

\newcommand{\bea}{\begin{eqnarray}}
\newcommand{\eea}{\end{eqnarray}}

\makeatletter
\def\slash{\@ifnextchar[{\fmsl@sh}{\fmsl@sh[0mu]}}
\def\fmsl@sh[#1]#2{%
  \mathchoice
    {\@fmsl@sh\displaystyle{#1}{#2}}%
    {\@fmsl@sh\textstyle{#1}{#2}}%
    {\@fmsl@sh\scriptstyle{#1}{#2}}%
    {\@fmsl@sh\scriptscriptstyle{#1}{#2}}}
\def\@fmsl@sh#1#2#3{\m@th\ooalign{$\hfil#1\mkern#2/\hfil$\crcr$#1#3$}}
\makeatother
\usepackage{color}

\definecolor{orange}{rgb}{0.9,0.2,0}
\definecolor{brown}{rgb}{0.7,0.3,0.2}
\definecolor{fuxia}{rgb}{1,0,1}
\definecolor{skyblue}{rgb}{0,0.1,0.9}
\definecolor{violetred}{rgb}{0.8,0.13,0.56}
\definecolor{deeppink}{rgb}{1.00,0.08,0.5}
\definecolor{pink}{rgb}{1.00,0.75,0.80}
\definecolor{orchid}{rgb}{0.85,0.44,0.84}
\definecolor{lightpink}{rgb}{1.00,0.71,0.76}
\definecolor{bluish}{rgb}{0,0.6,0.8}
\begin{document}

\title{Towards a formulation  of $f(R)$ supergravity }
\author{G. A.~\ Diamandis} \email{gdiam@phys.uoa.gr}
\author{A. B.~\ Lahanas} \email{alahanas@phys.uoa.gr}
\affiliation{University of Athens, Physics Department,\\
Nuclear and Particle Physics Section, GR--157 71  Athens, Greece}
\author{K. Tamvakis}
\email{tamvakis@uoi.gr}
\affiliation{University of Ioannina,  
Physics Department,
Section of Theoretical Physics, 
GR--451 10  Ioannina, Greece}
\vspace*{2cm}  
\begin{abstract}

The supersymmetric completion of
$R+R^2$ gravity is known to be equivalent to standard supergravity coupled to two chiral supermultiples with a K\"ahler potential of the no-scale type.  The generalization of this program to $f(R)$ gravity can be carried out in an analogous framework in terms of four chiral multiplets with appropriately chosen superpotential and K\"ahler potential.  Although the construction strategy has been known for sometime, we carry out this program further by setting these theories  in a closed form. The general framework considered can accommodate supergravity actions local in the scalar curvature, dual to ordinary $N = 1$ supergravities. Although these are in general plagued by ghosts, the framework studied in this work offers a possibility that  these  can decouple as can be seen in specific examples.

\end{abstract}
\maketitle



\section{Introduction}

Standard or minimal $N=1$ supergravity, being the supersymmetric completion of the Einstein theory, is of particular importance in particle physics, since it is considered to be the proper field theoretic limit of superstring theory at energies below the string scale. Furthermore, the supersymmetric completion of $R+R^2$ gravity turns out to be equivalent to standard supergravity coupled to two chiral supermultiplets with a K\"{a}hler potential of the no-scale type \cite{CECOTTI}. This poses immediately the more general question of the supersymmetric completion of an $f(R)$ gravity.
In the absence of supersymmetry $f(R)$ gravity theories are known to be equivalent to Einstein-Hilbert actions, if one introduces properly chosen auxiliary scalar fields. In doing that, the theory is first written in the Jordan-frame with the scalar curvature appearing linearly coupled to the auxiliary fields \cite{CAPO}. Under an appropriate Weyl transformation of the metric, that absorbs the auxiliary field dependent coefficient of the scalar curvature, the gravity sector becomes of the  Einstein-Hilbert form and the auxiliary fields become dynamical. This duality between a given $f(R)$ gravity and its Einstein form is not so easily implemented in the framework of a supergravity theory \cite{KETOV1,KETOV,SFERRARA,SFERRARA2}. Nevertheless, there is strong motivation for studying $f(R)$ supergravities based mainly on the fact that $R + R^2$ supergravities provide a fruitful framework to describe cosmological inflation \cite{ELLIS, KTAL}. In the simplest of these models, besides the graviton, the theory describes an additional scalar particle with mass squared dictated by the coefficient of the $R^2$ term.  This extra scalar  degree of freedom in  its dual description, in the Einstein frame, is the so-called inflaton field. The central appealing feature of the simplest quadratic curvature model, the Starobinsky model \cite{STAROB}, is that the predicted inflaton potential is suitable for slow-roll inflation to set in. The scale of inflation is set by the coupling of the $R^2$ term of the gravitational action. This single field inflaton model describes in an efficient way cosmological inflation in agreement with Planck and recent BICEP2 data \cite{PLANCK,BICEP,PLANCK2,BICEPPLANCK} . Nevertheless alternative or more general options may be open  that are  described by $f(R)$ theories.  A program for suitably embedding such models within supergravity, sometimes referred to as {\textit{supersymmetric completion}}, must necessarily start from a Jordan-frame supergravity action in which higher derivative terms are present. The equivalence, between the Jordan-frame description with auxiliary fields and the $f(R)$ description, apart from being technically involved, may not even be unique.

In the present article we undertake the task to formulate a supersymmetric completion of an $f(R)$ gravity theory. Our construction strategy consists in starting with the Jordan-frame theory with a number of chiral superfields coupled through a specifically chosen form of the kinetic function and the superpotential, which, nevertheless, is general enough to include classes of models. After integrating out the auxiliary degrees of freedom, gravity enters in the resulting theory through a general function of the curvature that, depending on the model, may or may not be local. Then, we proceed to study the corresponding Einstein-frame theory formulated in the standard way in terms of a number of chiral superfields coupled through a K\"{a}hler potential and a superpotential chosen as above. Thus, in this framework a supersymmetrization of an $f(R)$ theory is accomplished. The proposed scheme is general enough to include other partial approaches like chiral models. Nevertheless, unphysical degrees of freedom are in general present and their removal should be required. Although we have no general answer to this issue, we propose possible solutions in specific cases.

Our article is organized as follows. In the next section we review the standard formulation of supergravity in the Jordan-frame and its connection with the corresponding Einstein-frame formulation. In section III, assuming a specific but quite general form of the kinetic function and the superpotential, as well as a set of at least four chiral superfields, we proceed to derive the Jordan-frame supergravity and show the equivalence of its bosonic part to an $f(R)$-type of theory of gravity. In the same section we supply examples considering various classes of models and working out some of them. In section IV we show that the so-called chiral Lagrangian models are included as a special case in our scheme. In section V we analyze the formulation of the above theories in the Einstein frame. We discuss the, now, more transparent issue of ghost states and analyze classes of specific models. Finally, in the last section we briefly summarize our conclusions.

\section{$N=1$ supergravity }

In this section we review the standard formulation of $N=1$ supergravity in the Jordan-frame{\footnote{
We follow the notation and conventions of J. Bagger and J. Wess\cite{BAGGER}.}}. Although this is well-known material, its brief presentation will serve to establish notation as well as the general framework to be used for the subsequent constructions and models. We start with the action
\bea
{\cal{S}} \, = \, \int \, d^4 x \, d^2 \Theta \, 
2 \, {\cal{E}} \, \left(  
- \dfrac{1}{8} \, ( \overline{\cal{D}  }^2 - 8 \, {\cal{R}} \, ) \,  \Omega(\Phi, \bar{\Phi} )  +{\cal{ W}}(\Phi) 
\right) + (h.c.) 
\label{ccc}
\eea
in terms of a kinetic function $\Omega(\Phi,\overline{\Phi})$ and a superpotential ${\cal{W}}(\Phi)$. For simplicity no gauge multiplets are assumed.  In (\ref{ccc})  $ {\cal{E}}$  is the vierbein determinant  multiplet and  $ {\cal{R}} $ is the scalar curvature multiplet. Their bosonic parts are
\bea
{\cal{E}} \, & = & \,  \dfrac{e}{2} \, \left( 1 - \Theta^2 \,  \overline{M} \right) \nonumber \\
{\cal{R}} \, & = &  \,  - \dfrac{M}{6}  + \Theta^2 \, \left(  \dfrac{R}{12}  - \dfrac{ M \, \overline{M} }{9}  - \dfrac{b_\mu^2}{18}  
+ \dfrac{i }{6} \, D_\mu b^\mu \,   \right)  
\label{mults}
\eea
$M$ and $b_\mu$ are the auxiliary fields of the gravity multiplet.  Note that the chiral action in (\ref{ccc}) can also be expressed in an equivalent manner using the chiral multiplets and their kinetic multiplets as done by Cremmer et al \cite{CREMMER}. Whatever the method used,  when we expand the above action in components  we get  the Jordan form of the corresponding supergravity.  
For a number of chiral multiplets involved,  labeled by $ i =1,2, \cdots$, whose   scalar  and  auxiliary components  are $ \phi^i $ and $F^i$  respectively,  the bosonic  part of the supergravity Lagrange density, in the Jordan-frame, is given by

\bea
 e^{-1} \mathcal{L} &=&  \nonumber \\
&& \frac{\Omega}{6} \,  \left( \  R +  \frac{2}{3} M \bar{M} - \frac{2}{3} b_{\mu} b^{\mu}\right) - \Omega_{i \bar j} 
\nabla_{\mu} \phi^i \nabla^{\mu}{\bar \phi}^{\bar j} +  \Omega_{i \bar j} F^i {\overline F}^{\bar j}
 - \frac{i}{3} \left( \Omega_i \nabla_{\mu} \phi^{i}  - \Omega_{\bar i} \nabla_{\mu} {\bar \phi}^{\bar i} \right) b^{\mu}  \nonumber \\
& -&\frac{M}{3}  \, \left( \Omega_i \, F^i + 3 \, {\overline W} \right)  - \frac{\overline M}{3} \, \left( \Omega_{\bar i} \, {\overline F}^{\bar i} + 3 \, W \right) + W_i \, F^i + {\overline W}_{\bar i } \,  {\overline F}^{\bar i }
\label{Lag1}\,.
\eea
The usual strategy that is followed, in order to arrive at the well-known supergravity action in the {{Einstein frame}}, is to eliminate the auxiliary fields $M, b_\mu, F^i  $  by solving their corresponding equations of motion and subsequently performing a 
{\textit{Weyl transformation}} in order to pass to the Einstein frame in which the Lagrange density takes on the form 
$  e^{-1} \mathcal{L} = - R / 2 \, + \cdots$. 
Then, one gets the usual description of the $N=1$ supergravity described by a K\"ahler function given by 
\bea
{\cal{G} } \, = \, {\cal{K} } \, + \, \ln \,  { | {{W}} |  }^{2}\,.\, 
\label{kaler}
\eea
In this, the  {\textit{K\"{a}hler potential}} is given by
\bea
{\cal{K}} \, \equiv \, -3 \, \ln \, \left(  - \dfrac{ \Omega }{ 3} \right) \,.\, 
\label{KKK}
\eea
The elimination of the auxiliary fields is implemented in the following manner. The auxiliary field part of the Lagrangian can be written as 
\bea
e^{-1} \, {\cal L}_{aux} \, &=& \,  - \dfrac{\Omega}{9} \, b_\mu^2 \, + \, \dfrac{i}{9} \, b_\mu \,  \Omega \, (  {\cal K}_i \, \nabla^\mu \phi^i - h.c.     ) 
\nonumber \\
&& + \dfrac{\Omega}{9} \, {| \tilde{u}  |}^2 - ( \tilde{u} \, W + h.c. ) 
\nonumber \\
&&  - \dfrac{\Omega}{3} \,   {\cal K}_{i \bar{j} }  \, F^i \, {\overline F}^{\bar{j}} \, + \, ( F^i \, ( W_i +     {\cal K}_{i  } W ) + h.c )
\label{auxit}\,.
\eea
We have replaced $M$ with the combination $\tilde{u}$ defined as
$$
\tilde{u} \, \equiv  \, {\overline M} +  {\cal K}_{i  } \, F^i
$$
and $  {\cal K}_{i }  $ stands for $ \partial {\cal{K}} / \partial  \phi^i $.  In general, the subscript $i$ denotes differentiation with respect 
$ \phi^i $ while $ \bar{i} $ differentiation with respect  ${\bar \phi}^{\, \bar i}  $. 
Then, the equations for $b_\mu, \tilde{u}, F^i$ yield
\bea
&&b_\mu \,  \, = \, \,  \dfrac{i}{2} \,  (  {\cal K}_i \, \nabla^\mu \phi^i - h.c.     )
\label{bbb} \\
&&\tilde{u} \quad = \, \dfrac{9}{\Omega} \, {\overline W} 
\label{uuu} 
\\
&& {\cal K}_{i \bar{j} }  \, F^{i} \, =  \,  
 \dfrac{ 3 }{ \Omega } \,    ( {\bar W}_{\bar j} +     {\cal K}_{\bar{j}  } {\bar W} )
\label{fff} 
\eea
Provided that the K\"ahler metric   $  {\cal K}_{i \bar{j} }   $ has no zeroes and it is invertible, we may proceed denoting its inverse by 
$ {(K^{-1})}^{\overline{i} j}   $. Then, ({\ref{fff}}) gives
\be 
F^{i}\,=\,\dfrac{3}{\Omega} \,  ( \, \overline{W}_{\overline{j}}+{\cal{K}}_{\overline{j}}\overline{W} \, )
{(K^{-1})}^{\overline{j} i} 
{\label{ffff}}\,.
\ee
 
Thus, $\tilde{u}$ and $F^i$ are eliminated and plugging ({\ref{uuu}}) and (\ref{ffff}) into the Lagrangian (\ref{auxit}) one gets 

\bea
e^{-1} \, {\cal{L}}_{aux}^{F, \tilde{u}} \, = \,
\frac{3}{\Omega} \,
\left(
 \, D_{\overline{j}} \overline{W} \, {(K^{-1})}^{\overline{j} i} \, (D_i W) - \, 3 \, { | W | }^2
\right)
\label{Fu}
\eea
where 
$$ D_iW\,=\,W_i+{\cal{K}}_iW   $$
 is a covariant derivative and 
$ D_{\overline{i}}\overline{W} $ is  its conjugate. 
Notice that in (\ref{Fu}) we have not included as yet the result of the elimination of the field $b_\mu$. 
The elimination of the field $b_\mu$  is done in a trivial manner. Using  the solution for  $b_\mu$,  given in   (\ref{bbb}),  and plugging into (\ref{auxit})  we have terms  quadratic in the derivatives of the fields, which, when added to the kinetic terms already existing in (\ref{Lag1}), yield  the full Lagrangian
\bea
 e^{-1} \mathcal{L} = \,\frac{\Omega}{6} \, R
  - \Omega_{i \bar j} 
\nabla_{\mu} \phi^i \nabla^{\mu}{\bar \phi}^{\bar j} \, - \,  \dfrac{\Omega}{36} \, 
 { (  {\cal K}_i \, \nabla^\mu \phi^i - h.c.     ) }^2\,+\,\frac{3}{\Omega} \,
 \left(
D_{\overline{j}} {\overline{W}} \, {(K^{-1})}^{\overline{j} i} \, (D_i W) \, - \, 3 \, { | W | }^2
 \right)
 {\label{LANG}}\,.
\eea
This is the Jordan-frame Lagrangian.  
We can pass to the Einstein frame by performing a Weyl transformation given by 
\bea
e^m_\mu = e^{\prime \, m}_\mu \,  \Lambda 
\label{weyl}\,\,,
\eea
in which case the curvature term in the action gives, up to a total derivative,  
\bea
\dfrac{ e \, \Omega}{  6} \, R \, = \,  \dfrac{ e^\prime \, \Omega \,  \Lambda^2}{  6} \, ( \,  R' \, + \, 
6 \, \, g_{\mu \nu}^\prime \, \Lambda^{-1} \, \nabla_\mu^\prime \,  \nabla_\nu^\prime \, \Lambda \, )\,\,.
\eea
Taking 
\bea
\Lambda \, = \,  { \left(  - \dfrac{3 }{\Omega } \right) }^{1/2}\,=\,e^{{\cal{K}}/6}\,, 
\eea
after a partial integration, we arrive at 
\bea
\dfrac{ e \, \Omega}{  6} \, R \, = \,  - \dfrac{ e^\prime }{  2} \,  R^{\, \prime}  \, - \, 
\dfrac{3}{4} \,  e^\prime \, {g'}^{\mu\nu} \,  ( \nabla_\mu   ln \Omega ) \,  ( \nabla_\nu   ln \Omega )
 \,=\,e'\left\{\,-\frac{R'}{2}\,-\frac{1}{12}\left({\cal{K}}_i\nabla'\phi^i\,+\,{\cal{K}}_{\overline{i}}\nabla'\overline{\phi}^{\overline{i}}\right)^2\,\right\}\,. {\label{Einstein}}
\eea
On the other hand, the rest of the terms of $e^{-1}{\cal{L}}$ in ({\ref{LANG}}) become in the Einstein frame
\bea
e^{-1} \mathcal{L} \, = \, 
&-& \, {\cal{K}}_{i\bar{j}}\nabla'\phi^{i}\nabla'\bar{\phi}^{\overline{j}}\,+\,\frac{1}{12}\left({\cal{K}}_i\nabla'\phi^{i}-{\cal{K}}_{\bar{i}}\nabla'\bar{\phi}^{\bar{i}}\right)^2\,+\,\frac{1}{3}{\cal{K}}_i{\cal{K}}_{\bar{j}}\nabla'\phi^i\nabla'\bar{\phi}^{\bar{j}}\,
 \nonumber \\
&-& \, e^{{\cal{K}}} \,
\left(\,
D_{\overline{j}} {\overline{W}} \, {(K^{-1})}^{\overline{j} i} \, (D_i W) )\, - 3 |W|^2
\right)
\eea
Thus, dropping the prime, the  final bosonic Lagrangian, in the Einstein frame, takes on the form 
{\footnote{
Recall that we follow the notation of Bagger and Wess according to which the scalar curvature is opposite to that of other authors (f.e. Birrell's and Davis's). The metric signature is $ - + + + $. 
}}
\bea
e^{ -1} \, 
{\cal{L}}_{Einstein} 
\, = \,  - \frac{1}{2}R  - \,   {\cal{K}}_{i \bar{ j}}  \, \nabla_{\mu} \phi^i \nabla^{\mu}{\bar \phi}^{\bar j} 
\, - \,  e^{\, \cal{K}} \, 
( \, 
D_{\overline{j}} {\overline{W}} \, {(K^{-1})}^{\overline{j} i} \, (D_i W)   \, - \, 3 \, { | W | }^2
\, )\,.{\label{EIN}}
\eea
The dual description of the same action is obtained in an alternative manner to be described below.  Evidently both descriptions are equivalent.

\section{The dual description of $N = 1 $ supergravity} 

\subsection{Derivation of the dual $f(R)$ supergravity }

Having reviewed the salient features of the standard formulation of ordinary $N = 1$ supergravity we proceed to consider an alternative description that is capable of incorporating more general gravitational schemes, in which the gravity sector will not be of the Einstein-Hilbert form. As we have discussed in the introduction, the Jordan-frame theory, under an appropriate Weyl transformation of the metric, can be set in the  Einstein-Hilbert form, while the auxiliary fields become dynamical. However, a supersymmetric theory has its scalar field content in the form of chiral supermultiplets with interactions described by a K\"{a}hler potential and a superpotential. A particular case where this program is realized in a straightfoward way is the case of the supersymmetric completion of the $R + R^2$, implemented along the lines suggested in \cite{CECOTTI}, \cite{THEISEN}. To this end two chiral multiplets $\Phi,\,T$ are needed. We consider  a kinetic function $\Omega$ and a superpotential $W$ given by
\bea
&&
{\Omega } / {3}  \, = \,  -  \, ( \, T + \overline{T}  - \Phi  \overline{\Phi}  \, ) 
 \\
&& 
W \, \; \;  \, = \, 3\, M  \, \Phi \, ( T - 1/2 \, )   
\eea
Note that $\Omega\,=\,-3e^{-{\cal{K}}/3}$ gives 
\be {\cal{K}}\,=\,-3\ln(T+\overline{T}-\overline{\Phi}\Phi)\,.\ee
Then, we can immediately write down the supergravity action in the Einstein frame using ({\ref{EIN}}). 
In the direction  $\Phi = 0 $ 
\bea{
e^{ \, -1} \, {\cal{L}} \, = \, 
- \, \dfrac{R}{2} \, - \dfrac{ 3 \, { | \nabla_\mu \, T  |  }^2 }{   {( T + \bar{T}  )}^2  } \, 
- \, 3 \, M^2 \,   \dfrac{  \, { | T - 1/2 |  }^2 }{   {( T + \bar{T}  )}^2  }
}\,.
\eea
At $Im \, T = 0   $ and with $  \, Re \, T \, \equiv  \, \frac{1}{2} \, e^{ \sqrt{ \frac{2}{3}  }  \, \varphi }   $ 
this is brought to the form 
\bea
e^{ \, -1} \, {\cal{L}} \, = \, 
- \, \dfrac{R}{2} \, - \dfrac{1}{2} \, { ( \nabla_\mu \, \varphi   ) }^2  \, 
- \, \dfrac{ 3 \, M^2 }{ 4  } \, { \left( \, 1 - e^ { - \, \sqrt{2/3} \, \varphi  }  \right)   }^{\,2}
\eea
which is the celebrated Starobinsky's model, with $\varphi$ being the inflaton field and $M$ the scale of inflation. 
The dual description of the Starobinsky model follows in an alternative manner starting from the Jordan form (\ref{Lag1})  by eliminating the auxiliary fields $M, F_T,F_{\Phi}$ and $T$ (which, since it appears linearly in $\omega$, turns out not to have a kinetic term), except the  field $b_\mu$. The resulting theory is certainly equivalent (dual)  to the  ordinary supergravity theory but the field $b_\mu$ becomes dynamical in the new description. 
The resulting supergravity action is given by 
\bea
 e^{ \, -1} \, {\cal{L}}_{dual}  \, & = &\, \,   - \, \dfrac{R}{2} \, + \, \dfrac{R^2  }{  12 \, M^2 } \, 
 - \, \left(  \, \dfrac{\Phi \, \bar{\Phi} }{2} +  \dfrac{b_\mu^2 }{9 \, M^2}  \, \right) \, R \,  \nonumber \\
 &&- \, 3 \,  {|  \nabla_\mu \, \Phi \, |}^2 - i \, b_\mu \, ( \bar{\Phi} \,  \nabla_\mu \, \Phi - c.c \, ) \nonumber \\
 &&+ \, \dfrac{1}{3 \, M^2} \, { ( D_\mu \, b^\mu  ) }^2  \, +   \, \dfrac{b_\mu^4 }{27 \, M^2} 
 \, + \,   \dfrac{ b_\mu ^2}{3 } \,   + \,  \dfrac{ b_\mu ^2}{3 }  \, {\Phi \, \bar{\Phi} }  
 \label{staror}\,.
\eea
Note that in this formulation $ R ,  b_\mu^2 ,  b_\mu^4$ enter through the combination $ \overline{R}\,=\,R - 2 \, b_\mu^2 / 3  \, $. Note also that the  field $ b_\mu$ is dynamical. 
This is the  supersymmetric completion of the simple  Starobinsky model which includes, unavoidably, additional  terms.
However the physical content of both formulations is the same. In particular, this theory describes a graviton with two on-shell degrees of freedom and a real scalar field, the ``scalaron", encoded within the gravity sector, which is quadratic in the curvature, and in addition 
 the complex scalar of the $\Phi$ multiplet and a real scalar particle corresponding to the longitudinal component of $b_\mu$. The transverse components of $b_\mu$ decouple from the spectrum.  


Dualizing a general $f(R)$ gravity along the lines that led to (\ref{staror}) needs additional chiral multiplets. Here we follow the prescription given in \cite{CECOTTI} and define the real function $\Omega$ and the superpotential  $W$ as
\bea
\Omega &=&   T + \bar{T}  +  \left( Q \, \bar{\Phi}+ \Phi \, \bar{Q} \right) + \omega   \left( C, \bar{C}, \Phi, \bar{\Phi} , S, \bar{S} \right) 
\nonumber \\
W &=&   T \, \Phi + Q \, C    + h  \left( C, \Phi, S \right)
\label{cec1}
\eea
These include at least four chiral multiplets, namely $T, C, \Phi, Q$, but in order to cover more general cases we allow for additional multiplets denoted collectively by $S$.  Without loss of generality we have rescaled the multiplets $T,C,\Phi, Q$ accordingly, so that all relevant couplings other than those within $\omega$ or $h$  are set to unity. Note that both $T, Q$ multiplets appear linearly in the kinetic function and the superpotential.  The functions  $\omega$ and $h$ are for the moment arbitrary. They do not depend on either of the linearly coupled multiplets $Q$ and $T$. However they can depend on $C, \Phi$ as well as on the additional multiplets $S$, in general.  Note that the part  of $\Omega$ including $T +\overline{T}$ is reminiscent of the no-scale structure \cite{NOSCA} employed in dualizing the $R^2$ theory discussed above \cite{KTAL}. 

Before we proceed to describe how an $f(R)$ supergravity action can be constructed from this theory, it is important to point out a general property of the theories described by the functions (\ref{cec1}) above. Due to their linear dependence on $T$ these theories are invariant under holomorphic shifts of the modulus $T$
\bea
T \, \rightarrow \,T'\,=\, T + A(\Phi, C, S)\,\,.
\label{holom}
\eea
In (\ref{holom})  $A$ is an arbitrary holomorphic function.  These transformations leave the no-scale structure of $ {\Omega} $ intact.
Under this shift $\omega$ and $h$ undergo the following transformations
\bea
\omega  \,  &\rightarrow& \, \omega^\prime =  \, \omega \, + \,  A(\Phi, C, S) + {\bar{A}} (\bar{\Phi}, \bar{C}, \bar{S}) \nonumber \\
h  \,  &\rightarrow& \, h^\prime =   \, h  \, + \,   {\Phi} \, {A(\Phi, C, S)} \,
 \label{shift1}
\eea
Due to this property any part of  $\omega$  which is a sum of an analytic and an antianalytic function, say $ F + \bar{F} \, $, can be transferred to the superpotential as follows 
\bea
 h  \,  \rightarrow  \, h^\prime =  \, h  \, - \,  \Phi \, {F}
  \label{shift2}
\eea
or inversely, the part $h$ of the superpotential $W$ can be transferred to $\omega$ as 
\be
\begin{array}{l}
\omega   \,  \rightarrow  \, \omega^{''} =  \, \omega   \, - \,  \dfrac{h}{\Phi} -  \dfrac{\bar{h}}{\bar{\Phi}}\\
\,\\
h\,\rightarrow\,h''\,=\, 0
\end{array}
 \label{shift3}
\ee
Our task is facilitated if we opt to transfer any possible analytic and antianalytic part of $\omega$  to the superpotential. This makes simpler the treatment of the   K\"ahler metric to be used in the Einstein frame as in the  previous section.  Therefore, unless otherwise stated, the $\omega$ is assumed not to include a sum of analytic and antianalytic functions. 

The methodology to pass from the  Jordan-frame form of this supergravity to its $f(R)$ description is to eliminate all auxiliary fields except the vector field $b_\mu$ which becomes dynamical. This is exactly what  was done in the $R^2$ theory studied before. 
When this is  done, the Lagrangian attains its {\textit{dual form}} 
 \bea
 {\cal{L}} \left(   R - \dfrac{2}{3} \, b_\mu^2 \, , D_\mu b^\mu, \cdots    \right)\,\,. 
 \label{dual1}
 \eea
 This depends on the scalar curvature and $b_\mu$ through the combination $R-2 \, b_{\mu}^2 / 3$. The reason for it will be  explained shortly. Note that  (\ref{dual1}) depends also on $D_\mu b^\mu$ so that $b_\mu$ is dynamical. 
 The ellipses in (\ref{dual1}) denote additional terms  which depend on the specifics of the models under study.

 In the following we  describe in detail how the elimination of auxiliary fields, leading to the Lagrangian  (\ref{dual1}),   is implemented.  
 The fields $T$ and $Q$ enter linearly in the kinetic function, the former having no kinetic term and the latter having only a mixed term
\bea
\int\,d^4x\,\sqrt{-g}\left\{-(\nabla_{\mu}q)(\nabla^{\mu}\overline{\phi})-h.c.\right\}\,=\,\int\,d^4x\,\sqrt{-g}\left\{\,q\,\Box\overline{\phi}\,+h.c.\,\right\}\,.
\eea
Thus, the fields $\tau,\,q$ are purely auxiliary. We therefore, proceed to consider the minimization of the action with respect to them giving their equations of motion, which will be just constraint  equations. They are
\begin{subequations}
\begin{align}
&\dfrac{\delta{\cal{S}}}{\delta\tau}\, =\,\dfrac{1}{6}\left( R + \frac{2}{3} \, \overline{M} M -\frac{2}{3} \, b_\mu^2 \right)-\phi \, \overline{M}+F_{\phi} + \dfrac{i}{3} D_{\mu} b^{\mu} \,=\,0 \label{24a} \\
&\dfrac{\delta{\cal{S}}}{\delta F^{\tau}}\, =\,\phi\,-\dfrac{M}{3}\,=\,0 \label{24b}  \\
&\dfrac{\delta{\cal{S}}}{\delta q}\, =\, \dfrac{i}{3}  \, \overline{\phi}  \,  D_{\mu} b^{\mu}  + 
\dfrac{2 \, i}{3}  \,  b^{\mu}  \, D_\mu \overline{\phi}  \, +
\dfrac{\overline{\phi}}{6} \left( R+\frac{2}{3}\overline{M}M-\frac{2}{3} b_\mu^2 \right) +
\Box \, \overline{\phi} - c \, \overline{M} + F^c - \dfrac{\overline{M}}{3} \, {\overline F}^{\bar{\phi}}\,=\,0 \label{24c}  \\
&\dfrac{\delta{\cal{S}}}{\delta F^q}\, =\, \overline{F}^{\bar{\phi}} + c-\dfrac{M}{3}\, \overline{\phi}\,=\,0
\label{24d} 
\end{align}
\end{subequations}
Note that these are independent of the $\omega$ and $h$ functions that complete  the real function $\Omega$ and the superpotential $W$ in Eq. (\ref{cec1}). Solving them allows us to determine $M, c, F^\phi$ and $F^c$. In particular, using  (\ref{24b}) one expresses $M$ in terms of $\phi$ and then   (\ref{24a}) is solved to express $F^\phi$ in terms of other fields as well. Then, (\ref{24d}) can be solved to find $c$. Lastly, (\ref{24c}) is solved to yield the auxiliary field $F^c \,$. The resulting solutions are 
\begin{subequations}
\begin{align}
&M \, = \, 3 \, \phi   \label{25a} \\
&F^\phi = - \, \dfrac{1}{6} \,    \left(   \,  R - \dfrac{ 2 \, b_\mu^2 }{ 3 }     \,   \right) 
+ {2 \,  |  \phi | }^2 \, - \, \dfrac{i}{3} \, D_\mu  b^\mu      \label{25b}  \\
&c \,  = + \,  \dfrac{1}{6} \,  \left(   \,  R - \dfrac{ 2 \,b_\mu^2 }{ 3 }     \,   \right) 
- { \,  |  \phi | }^2 \, - \, \dfrac{i}{3} \, D_\mu \, b^\mu      \label{25c}  \\
&F^c \, = \, - \square \, \bar{\phi} +  \dfrac{\bar{\phi} }{6} \,  \left(   \,  R - \dfrac{ 2 \,b_\mu^2 }{ 3 }     \,   \right) 
\, - \, 2 \, \bar{\phi} \,  { |  \phi | }^2 -  i \, \bar{\phi}  \, D_\mu \, b^\mu  - \dfrac{2 \, i}{3}  \, b^\mu  \,  \nabla_\mu \bar{\phi}
 \label{25d}
\end{align}
\end{subequations}
In addition to (\ref{25a}) - (\ref{25d}), we  have the equations of motion for the remaining auxiliaries $M, \, F^{i}$ with $i=\phi,c,s$, namely $\dfrac{\delta{\cal{S}}}{\delta M}\,=\,\dfrac{\delta{\cal{S}}}{\delta F^{i}} \,=\,0$. These equations determine $F^s , F^{\tau},\,F^q,\,q$, depending on the details of the model as encoded in  $\omega , h$. They are  
\begin{subequations}
\begin{align}
&\dfrac{\delta{\cal{S}}}{\delta F^i} \, = \,
\Omega_{ i \bar{j} } \, {\overline{F}}^{\bar{j}} \, - \, \dfrac{\Omega_i}{3} \, M + W_i \, = \, 0 
\quad , \; \; \text{for} \quad i = \phi,  \, c,  \, s  \label{26a} \\
&\dfrac{\delta{\cal{S}}}{\delta M} \, \, \, = \,
\dfrac{\Omega}{9} \,   \overline{M}- \, \dfrac{\Omega_i}{3} \, F^i - \overline{W} \, = \, 0 \end{align}
\label{26b}
\end{subequations}
Note that Eq. (\ref{26a}) for $ i = s$ yields
\bea
{\overline{F}}^{\bar{s}} \, = \,  \omega_{s \bar{s}}^{-1} \, \left(  \phi \, \omega_s - h_s - \omega_{s \bar{\phi}} \, 
{\overline{F}}^{\bar{\phi }} - \omega_{s \bar{c}} \, 
{\overline{F}}^{\bar{c}} \right)\,, 
\label{fss}
\eea
where we have tacitly assumed that $  \omega_{s \bar{s}}  \neq 0$. 
Using (\ref{25a}) - (\ref{25d}) we may express $F^s$ in terms of $\phi, b_\mu, R$ and $s$.  The rest will yield $q, F^q, F^\tau$ but these are actually redundant, as we  see shortly. Nevertheless, for reasons of completeness we may present that $ \delta S / \delta F^c =   \delta S / \delta F^{\phi} = \delta S / \delta M   = 0   $ yield respectively, 
\begin{subequations}
\begin{align}
&q=-h_c+ \phi\omega_c-\omega_{c\bar{c}}\overline{F}^{\bar{c}} - \omega_{c\bar{\phi}}\overline{F}^{\bar{\phi}}
- \omega_{c\bar{s}}\overline{F}^{\bar{s}}   \label{28a} \\
&{\overline F}^{\bar q} = - {\tau} -  {h}_{{\phi}} + {\phi} \, ( \bar{q} + \omega_{{\phi}} )
- \omega_{\phi \bar{c}}\overline{F}^{\bar{c}}-\omega_{\phi \bar{\phi}}\overline{F}^{\bar{\phi}} -
\omega_{\phi \bar{s}} {\overline F}^{\bar s} 
  \label{28b} \\
&F^{\tau}={\overline \phi} \,  \Omega - 3 \, \overline{W} - {\overline \phi} \, F^q - {\overline q} \, F^{\phi}  - 
\omega_{\phi} \, F^\phi - \omega_{c} \, F^c - \omega_{s} \, F^s
 \label{28c} 
\end{align}
\end{subequations}
In  deriving (\ref{fss}) and (\ref{28a}) - (\ref{28c}) we have used Eq. (\ref{25a}), i.e.    
$ M = 3 \, \phi$. 

Since the fields $M, c,F^c ,F^\phi$  have all been expressed in terms of $(\,\phi,\, b_{\mu},\,R\,)$ and $F^s$ is expressed in terms of $(\,\phi,\, b_{\mu},\,R,\,s\,)$, it follows that  $q,\,F^q,\,F^\tau$ are all expressed in terms of $(\,\phi,\, b_{\mu},\,R,\,s\,)$  as well. 
These can be replaced back into the Lagrangian (\ref{Lag1}) to get the final result. However, as  already pointed out, there is much simplification in implementing this and  Eqs. (\ref{28a}) - (\ref{28c}) are actually redundant.  The important point is the fact that the Lagrangian (\ref{Lag1}) is linear in the fields $\tau, q, F^\tau, F^q$.  As a consequence, it can be written as
\bea
e^{-1} \, {\cal{L}} \, = \left(  \, \tau \, \dfrac{\delta S}{\delta \tau} + F^\tau \, \dfrac{\delta S}{\delta F^\tau} +  q \, \dfrac{\delta S}{\delta q} + F^q \, \dfrac{\delta S}{\delta F^q } +  h.c. \right) \, + \, 
e^{-1} \, {\cal{L}}( \tau=q=F^\tau=F^q=0 ) 
\eea 
In this, the last part denotes  the Lagrangian (\ref{Lag1}) with $  \tau, F^\tau, q, F^q $   set to zero. Thus, using the Eqs. 
(\ref{24a})- (\ref{24d}),  or equivalently (\ref{25a}) - (\ref{25d}), we arrive at 
\bea
e^{-1} \, {\cal{L}}_0 \, \equiv e^{-1} \, {\cal{L}}( \tau=q=F^\tau=F^q=0 ) 
\label{Lag3}
\eea 
which, using (\ref{Lag1}) , can be cast in the form
\bea
 e^{-1} \, {\cal{L}}_0 \, &=& \,   
   \dfrac{\omega}{6}   \left( \  R +  \frac{2}{3} M \bar{M} - \frac{2}{3} b_{\mu} b^{\mu}\right) 
-  \omega_{a \bar b}  \nabla_{\mu} \phi^a \nabla^{\mu}{\bar \phi}^{\bar b} +  \omega_{a \bar b} F^a F^{\bar b} - \frac{i}{3} \left( \omega_a \nabla_{\mu} \phi^{a} 
 - \omega_{\bar a} \nabla_{\mu} {\bar \phi}^{\bar a} \right) b^{\mu}  \nonumber \\
&& - \frac{M}{3}  \, \left( \omega_a F^a + 3 \bar h \right)  - \frac{ \bar M }{3}\left( \omega_{\bar a} {\overline F}^{\bar a} + 3 h \right) + 
h_a F^a + {\bar h} _{\bar a}  {\overline F}^{\bar a} \quad .
\label{Lagfin}
\eea 
{\textit{This is the dual description of the ordinary $N = 1 $  supergravity having the general form (\ref{dual1}). It is important to point out that in (\ref{Lagfin})  the indices $a, b$ run over $\phi, c, s$ only. The fields $M, c, F^c, F^\phi$ and $F^s$ are given by (\ref{25a}) - (\ref{25d}) and (\ref{fss}) respectively}}
\footnote{
In the absence of  additional  fields $S$,    Eq. (\ref{Lagfin}) still holds true by taking $F^s=0$.}.

Thus, as already pointed out,  (\ref{28a}) - (\ref{28c}) are not actually needed.  
Upon replacing $M, c, F^c, F^\phi$ and $F^s$, the derived Lagrangian (\ref{Lagfin}) depends on $\phi, s, b_\mu$ and the curvature 
$R$.  Its specific form depends on the details encoded in $\omega$ and $h$ functions  
 defined by (\ref{cec1}). The derivation of  Lagrangian (\ref{Lagfin}) is an important result since through this we can  express in closed form, given the functions $\omega$ and $h$, the dual form of an ordinary  supergravity theory. 
Specific examples will be given in the next section.  

For the sake of completeness, as a simple exercise, we show that Lagrangians with $\omega$ and $h$ related by a shift  transformation, as in (\ref{shift1}), yield the same theory. In particular, up to a divergence, the difference of the two Lagrangians,  defined  respectively by 
$\omega^\prime, h^\prime$ and  $\omega , h$ ,  is found, in a straightforward manner,  to be  
\bea
e^{-1} \, {\cal{L}}_0^\prime   -  e^{-1} \, {\cal{L}}_0   \,   \, = \, A  \, 
\left[ \, F^\phi  + \, \dfrac{1}{6} \,    \left(   \,  R - \dfrac{ 2 \, b_\mu^2 }{ 3 }     \,   \right) 
- {2 \,  |  \phi | }^2 \, + \, \dfrac{i}{3} \, D_\mu  b^\mu  \,  \right] +  h.c.\,,
\eea
which vanishes by Eq. (\ref{25b}).

It is important to note that in order to obtain a local action, i.e. one without derivatives of the curvature $R$, we have to restrict appropriately  the function $\omega$. In fact, derivatives of the curvature $R$ arise from derivatives of the field $c$, which is given by (\ref{25c}),  and therefore potential  sources of $\partial R$ are  the kinetic terms and the coupling of $b_\mu$ to the current $ \sim i (  \omega_a \, \nabla_\mu \phi^a - h.c. ) $.  In the case that the function $\omega$ involves $ c \, \bar{c}  $ mixings, unavoidably the theory will be nonlocal. 
However the requirement of locality restricts even further the possible form of the function $\omega$.  We can isolate the terms in (\ref{Lagfin}) that include derivatives of the field $c$ and give  rise to $\partial R$ . These are given by 
\bea
 e^{-1} \, {\cal{L}}_{\partial R}  \, &=& \,   
-  \omega_{c \bar c}  \, \nabla_{\mu} c \,  \nabla^{\mu}{\bar c} -   
\left( \,  \sum_{\bar{b}  \neq \bar{c} } \omega_{c \bar b}  \, \nabla_{\mu} c \, \nabla^{\mu}{\bar \phi}^{\bar b} \, + \, h.c. \, \right)
 - \frac{i}{3} \left( \omega_c \nabla_{\mu} c - h.c  \right) \, b^{\mu}  \nonumber \\
 && 
\,=\, -  \omega_{c \bar c}  \, \nabla_{\mu} c \,  \nabla^{\mu}{\bar c} +   
\left( \, c \,  \sum_{\bar{b}  \neq \bar{c} } \nabla_{\mu}  ( \omega_{c \bar b}   \, \nabla^{\mu}{\bar \phi}^{\bar b} ) \, + \, h.c. \, \right)
 + \frac{i}{3} \, \left(  \, c \,   \nabla_{\mu} ( \omega_c \, b^{\mu} )   - h.c  \right) \, 
 \label{nonloc}
\eea 
where the second line follows by partial integrations. From the last term, including the field $b^\mu$, we see that absence of 
$\partial R$ terms  requires that $\omega_c$ is independent of the field $c$, i.e. the function $\omega$ is linear in the field $c$, and hence $\bar{c}$. This in turn implies that $\omega_{c \bar b}$ is independent of $c$ too,  and therefore the middle term in (\ref{nonloc})  does not give rise to nonlocal terms either. Lastly, due to its linearity in $c$, the function $  \omega_{c \bar c}  $ vanishes and therefore the first term in  (\ref{nonloc})   does not produce nonlocal terms either. Therefore we conclude that the most general form of 
$\omega$ ensuring absence of nonlocal terms is
\bea
\omega \, = \, c \, f ( { \phi}^{ b} , {\bar \phi}^{\bar b} ) +  \xi( { \phi}^{ b} , {\bar \phi}^{\bar b} ) \, + \, 
h.c. \, \quad \quad  \quad (\phi^b \neq c)
\label{omloc}
\eea
The  functions $f , \xi$ are in general  arbitrary functions of the fields ${ \phi}^{ b} , {\bar \phi}^{\bar b}   $. Note that without loss of generality we absorb the analytic and antianalytic part of $\omega$ in the superpotential [ see discussion following Eq. (\ref{shift3})~], and on these grounds the function $f $ in (\ref{omloc})  should  necessarily  depend on $ {\bar \phi}^{\bar b}  $ while $\xi$ should depend on both $  { \phi}^{ b} , {\bar \phi}^{\bar b} $. 

\subsection{Specific models}

In order to see how (\ref{Lagfin}) can be implemented in describing  $f(R)$ supergravity theories we study some specific examples given below. Note that the Lagrangian of Eq. (\ref{Lagfin}) has a very simple form for vanishing $\omega$. Therefore, we start by studying such models. 

\subsubsection{A minimal model}
By the term minimal we  refer  to a  model with a vanishing $\omega$, no additional chiral multiplets $S$ and a special $h$ defined by 
\bea
\omega \, = \, 0 \quad , \quad h(\Phi, C) \, = \,  - \Phi \, F(C)\,
\label{we11}
\eea
corresponding to a superpotential linear in $\Phi$.  Then, from (\ref{Lagfin}) we get, 
\bea
 e^{-1} \, {\cal{L}}_0 \, =  -   {\overline M} \, h +   h_\phi F^\phi + h_c \, F^c + h.c.   
 \label{Lagex1}
\eea
Replacing $M, c, F^\phi, F^c$ in this, using (\ref{25a} - \ref{25d}), we get in a straightforward manner 
\begin{eqnarray}
 e^{-1} \, {\cal{L}} \, = && \, \left( \dfrac{R}{6} +  { | \phi |  }^{2}  - \dfrac{b_\mu^2 }{9} \, 
 + \dfrac{i}{3} \, D_\mu b^\mu \,  \right) \, F(c) \,  + \, 
\, \Bigg(  \,  \phi \, \Box \, \bar{\phi}   \,   + \,    2  { | \phi |  }^{4} - \dfrac{R}{6} \, { | \phi |  }^{2}         
\nonumber  \\
&& 
+  \,   i \,  D^\mu b_\mu \, \, { | \phi |  }^{2} 
+ \, \dfrac{b_\mu^2}{9} \,  { | \phi |  }^{2} \,
+  \dfrac{2 i}{3} \, b^\mu \, {\phi} \, \nabla_\mu {\bar \phi }
  \Bigg) \, F^\prime(c)  
\, + \, h.c 
\label{mainLall}
\end{eqnarray}
where $c$ is the scalar component of the chiral field $C$ given in (\ref{25c}) as $c \,  = \,R / \, 6 - {b_\mu}^2 / \, 9 
- |\phi |^2 \, - \, i \, D_\mu \, b^\mu / \, 3 $. 
If for simplicity  we take the arbitrary function $F(C)$ real and collect the terms that depend solely on the curvature $R$, we get
\bea
e^{-1} \, {\cal{L}} \, =  \, 
\dfrac{R}{3} \, F(R / 6 ) \, + \, \cdots\,\equiv\,f(R)\,+\,\dots
\label{frmin2}
\eea
Therefore this theory indeed describes an $f(R)$ supergravity with $f(R) =  \dfrac{R}{3} \, F(R / 6 ) $.  Thus, an $f(R)$ gravity can be embedded in the particular  supergravity theory defined by the above $\Omega=T+\overline{T}+Q\overline{\Phi}+\overline{Q}\Phi$ and $W=T\Phi+QC-\Phi F(C)$,  provided the function $F(C)$ is related to $f(R)$ as in ({\ref{frmin2}}).

\subsubsection{Next to minimal models}
We can easily generalize the previous case by considering 
\bea
\omega \, = \, 0  , \,\,\,h(\Phi, C) \, \equiv \,  - P(  \Phi , C)
\label{we3}
\eea
i.e. $\omega$ is still taken vanishing and $h$ is a general function of $C, \Phi$,  not necessarily linear in $\Phi$.  As in the previous model no additional fields $S$  are present. In this case too, the Lagrangian is given by  (\ref{Lagex1}). It is found, in a straightforward manner, that this case leads to  the dual theory  given by
\begin{eqnarray}
 {\color{black} e^{-1} \, {\cal{L}}_B  \,   } & {\color{black}  = }& \, {\color{black}  \,  \left( \dfrac{R}{6} +  { | \phi |  }^{2}  - \dfrac{b_\mu^2 }{9} \, 
 + \dfrac{i}{3} \, D_\mu b^\mu \,  \right) \, H(c, \phi) \,  + \, 
\, \Bigg(  \,  \phi \, \square \, \bar{\phi}   \,   + \,    2  { | \phi |  }^{4} - \dfrac{R}{6} \, { | \phi |  }^{2}     }
\nonumber  \\
&& 
{\color{black} 
+ \, \,   \,   i \,  D^\mu b_\mu \, \, { | \phi |  }^{2} 
+ \, \dfrac{b_\mu^2}{9} \,  { | \phi |  }^{2} \,
+  \dfrac{2 i}{3} \, b^\mu \, {\phi} \, \nabla_\mu {\bar \phi }
  \Bigg) \,   \dfrac{ \partial \, H(c, \phi) }{ \partial \, c } 
\,      }
\nonumber  \\
&&
  \,  + \, \left( \dfrac{R}{6} -   \dfrac{b_\mu^2 }{9}  -2  { | \phi |  }^{2} \, 
 + \dfrac{i}{3} \, D_\mu b^\mu \,  \right) \, \phi \,  \dfrac{ \partial \, H(c, \phi) }{ \partial \, \phi }     
\, + \, h.c.
\label{gen}
\end{eqnarray}
$c$ is again given by (\ref{25c}) and $ H(c, \phi)  \equiv P(c, \phi  ) / \phi $. 

As an application of it, and in order to establish connection with previous works,  consider the case where  the function $P \equiv - h $ depends only on $\Phi$. Then the middle term in (\ref{gen}) vanishes and by a straightforward calculation it is found that the Lagrangian (\ref{gen})  takes on the form 
\begin{eqnarray*}
  e^{-1} \, {\cal{L}}_B  \,    & = & \, 
 - \, 3 \, ( \Phi \, \overline{ h } + \overline{\Phi} \, { h } ) - \left( \, \dfrac{R}{3}  - \dfrac{2 \, b_\mu^2  }{9}  - 4 \,  {| \phi |  }^{2}  \right) \, 
 Reh^\prime
 \, - \, \dfrac{2}{3} \, b^\mu \, \nabla_\mu Im h^\prime
\end{eqnarray*}
In this primes denote derivatives with respect $\Phi$ . Also 
in deriving this  derivatives of $b_\mu$ have been transferred to derivatives of the imaginary part $ Im h^\prime $ by partial integrations. This facilitates a great deal since in this way  the field $b_\mu$ is manifestly auxiliary with equation of motion given by
\begin{eqnarray*}
b_\mu \, = \,  \dfrac{3}{2} \,  \dfrac{  \nabla_\mu Im h^\prime }{ Re h^\prime }
\label{eqbmu}
\end{eqnarray*}
which when plugged into the above  Lagrangian  leads to 
\begin{eqnarray*}
  e^{-1} \, {\cal{L}}_B  \,    & = & \, 
 - \, 3 \, ( \Phi \, \overline{ h } + \overline{\Phi} \, { h } ) -  
 \, Reh^\prime  \, \left( \, \dfrac{R}{3}  - 4 \,  {| \phi |  }^{2}  \right) \, 
 - \, \dfrac{1}{2} \, \dfrac{  { ( \nabla_\mu Im h^\prime ) }^2 }{  Re h^\prime } 
\end{eqnarray*}
This, modulo rescalings of fields  is identical to Eq (3.16)  of  \cite{SFERRARA2}. By a Weyl rescaling this is brought to the Einstein form
\begin{eqnarray*}
  e^{-1} \, {\cal{L}}_B  \,    & = & \, 
 - \dfrac{R}{2} \, - \, \dfrac{3}{4} \, {\left( \dfrac{  {  \nabla_\mu Re h^\prime  } }{  \, Re h^\prime } \right) }^2 \, 
 - \, \dfrac{3}{4} \, {\left( \dfrac{  {  \nabla_\mu Im h^\prime  } }{  \, Re h^\prime } \right) }^2 \,
 - \dfrac{27}{4} \, \dfrac{( \phi \, \overline{ h } + \overline{\phi} \, { h } ) ) }{ { ( Re h^\prime ) }^2} \, 
 + \, \dfrac{ 9  \, {| \phi |  }^{2} }{  Re h^\prime}
\end{eqnarray*}
In this way we have reproduced, using the formalism presented in this work,  the action given in  \cite{SFERRARA2}, which,  contrary to earlier claims,   is not a $R+R^2$ theory.

\subsubsection{Models with nonzero $\omega$}

We can generalize the previously considered model by allowing $ \omega \neq 0$.  Consider for instance, 
\bea
\omega \, = \, - \lambda \, { |  \Phi | }^4 - k \, { |  \Phi | }^2  \quad , \quad 
h  \, \equiv  \, - P(C, \Phi )  
\label{we44}
\eea
This model, as any model having $ \omega \neq 0$,  cannot be represented by the chiral actions  studied in  \cite{KETOV}, which we shall  discuss in the following section.  
Notice that in (\ref{Lagfin}) the contributions of $\omega$ and $h$ are additive.  Due to this, 
the contribution of $\, \omega = -  \lambda \, { |  \Phi | }^4 - k \, { |  \Phi | }^2 $  is added to the Lagrangian 
(\ref{gen}) found before. Thus, from the $\omega$ dependent terms in (\ref{Lagfin}),  it is found 
by a straightforward calculation, that the additional terms are given by
\bea
e^{-1} \, \Delta {\cal{L}} \, & =   &  \, 
k \, \Bigg(
- \dfrac{1}{36} \,  {\left( R -   \dfrac{2 \, b_\mu^2 }{3}  \right) }^{ 2} + \dfrac{ \, { |\, \phi |  \,}^{2} }{6} \, 
{\left( R -   \dfrac{2 \, b_\mu^2 }{3}  \right) }  \, - \, \dfrac{1}{9} \,  { \left( D_\mu b^\mu \right) }^{\,2} \,
\nonumber \\
&&
 - \dfrac{ i}{3} \, b^\mu \, ( \, {\phi} \, \nabla_\mu {\bar \phi } - c.c. \, ) 
 \, +   \, { |  \nabla_\mu \, \phi \, |}^2 \, - \, { | \, \phi \, |}^4
 \Bigg)
 \nonumber \\
 &&
 + \, \lambda \,  { | \, \phi \, |}^2 \,   
 \Bigg( 
 - \dfrac{1}{9} \,  {\left( R -   \dfrac{2 \, b_\mu^2 }{3}  \right) }^{ 2}
 \, + \, \dfrac{11 \, {| \, \phi \, |}^2}{6 \,} \,  {\left( R -   \dfrac{2 \, b_\mu^2 }{3}  \right) } \, 
  \, - \, \dfrac{4}{9} \,  { \left( D_\mu b^\mu \right) }^{\,2} \, 
 \nonumber \\
 &&
 - \dfrac{2 \,  i}{3} \, b^\mu \, ( \, {\phi} \, \nabla_\mu {\bar \phi } - c.c. \, )
 \, +   \, 4 \, { |  \nabla_\mu \, \phi \, |}^2 \, - \, 9 \, { | \, \phi \, |}^4
 \Bigg)
 \label{gen2}
 \eea
Note that all models studied so far  are  in closed form and also local, i.e. no derivatives of the curvature are present.  The latter is related to the fact that in the examples studied in this section, the $\omega$ functions are either vanishing or have no dependence on $C, \overline{C}$ at all. These models in their standard  description  of $N=1$ supergravity, as we  discuss later, have in general ghost states which should decouple for them to be well defined. 


\section{Relation to  chiral Lagrangians }

This section is devoted to finding  the correspondence of the theories derived here and given by (\ref{Lagfin}) and the chiral theories studied in \cite{KETOV}.  We  argue that  the Lagrangian (\ref{Lagfin}), for properly chosen $\omega$ and $h$ functions, yields the chiral actions studied in \cite{KETOV} and, therefore, they are of broader applicability.  The aforementioned chiral actions have the general form  
\bea
S \, = \, \int \, d x^4 \, d^2 \, \Theta \, \, 2 \, {\cal{E}} \,    {\cal{F}} ({ \cal{R}},  T( {\cal{R}}  )  )     \, + \,  h.c\,\,\,. 
\label{chiral}
\eea
In (\ref{chiral})  the bosonic parts of the vierbein determinant $ {\cal{E}}$  and the scalar curvature $ {\cal{R}} $ multiplets are given by (\ref{mults}). 
The kinetic multiplet of $ \cal{R} $, denoted by $T(  \cal{R} )  $,  has as scalar component the $ \Theta^2 $ component of $ \cal{R} $ and is given by
\bea
{T(  \cal{R} )}  \, &=& \, - \dfrac{R}{3} + \dfrac{2}{9} \, M  \overline{M} + \dfrac{2}{9} \, b_\mu^2 + \dfrac{2 i }{3} \, D_\mu b^\mu \,
\nonumber \\
&+& \, \Theta^2 \, \left( - \dfrac{ R }{9} \,  \overline{M}   +  \dfrac{2}{3} \, \Box \, \overline{M}  + \dfrac{4}{27} \,  \overline{M}  \, { | M | }^{\,2} 
+ \dfrac{2}{27} \,  \overline{M}  \, b_\mu^2 +  \dfrac{2 i }{3} \, \overline{M} \, D_\mu b^\mu \, + \, 
 \dfrac{4 i }{9} \, b^\mu \,   D_\mu  \overline{M} \,
\right)
\eea
This may differ by an overall multiplicative factor from the one used by other authors. Here we follow the notation of Wess and Bagger. 
The chiral action (\ref{chiral})  corresponds  to a  supergravity Lagrangian with $\omega$ and $h$ given by 
\bea
\omega \, = \,0 , \quad h ( \Phi, C )  \equiv    {\cal{F}} ( -  \Phi / 2,  - 2 \, C   )  )  
\label{hhh}
\eea
or, due to the shift property,  to a Lagrangian  with
\bea
\omega \, = \, -   {\cal{F}} ( - \Phi / 2,  - 2 \, C   )  / \Phi  + h.c. \, , \quad h = 0 \,\,\,.
 \label{ooo}
\eea
Therefore, one can have three equivalent constructions of one and the same theory.  
The Lagrangians of Eq. (\ref{chiral}) are local in the sense that no derivatives of $R$  are present. 

In order to see how chiral actions of the form (\ref{chiral}) can be used to supersymmetrize an $f(R)$ gravity, consider an arbitrary power $ {T(  \cal{R} )}^{\, m}  $. This is of the form
\bea
{T(  \cal{R} )}^{\, m}  \, = \, {\left(   - \dfrac{R}{3} \right)}^{\, m} \, + \cdots
\eea
where we have isolated the highest power of the curvature appearing in the first component of this multiplet.  
Ellipses denote additional terms that are not shown. From this we get
\bea
{\cal{R}} \, {T(  \cal{R} )}^{\, m}  \, = \, \Theta^2 \, \dfrac{R}{12} \,  {\left(   - \dfrac{R}{3} \right)}^{\, m} 
+ \cdots
\eea
and therefore
\bea
2 \, { \cal{E}} \, {\cal{R}} \, {T(  \cal{R} )}^{\, m} \biggr\rvert_{ \Theta \Theta}  \, + \, ( h.c. ) =  
e \, \dfrac{ R^{\, m+1}}{ 6 \, {( -3 )}^{\,m} } \, + \, \cdots
\eea
Using this, we get
\bea
\int \,  d^2 \, \Theta \, \, 2 \, {\cal{E}} \,  \left( \,  \sum_m \, a_m \,   {\cal{R}} \, {T(  \cal{R} )}^{\, m}   \, \right) \, + \,  h.c. \, = \, 
 e \, \sum_m \, a_m \,    \dfrac{ R^{\, m+1}}{ 6 \, {( -3 )}^{\,m} } \, + \, \cdots
 \label{Fall}
\eea 
The right-hand side of this describes an $f(R)$ gravity theory given by 
\bea
f(R) \, = \, \sum_m \, b_m \, R^{\, m+1} \quad  \;  \;  \text{with} \quad b_m =  \dfrac{ a_m }{ 6 \, {( -3 )}^{\,m} } \,\,.
\eea
Therefore on account of Eq (\ref{Fall}) we conclude, with the given definitions for $ {\cal{R}}$ and ${T(  \cal{R} )}$, that  if
\bea
 {\cal{F}} ({ \cal{R}},  T( {\cal{R}}  )  ) \, \equiv \, 6 \,  \sum_m \, b_m \,  {( -3 )}^{\,m}   \,  {\cal{R}} \, {T(  \cal{R} )}^{\, m}\,, 
 \label{kkk}
\eea
then 
\bea
\int \,  \, d^2 \, \Theta \, \, 2 \, {\cal{E}} \,    {\cal{F}} ({ \cal{R}},  T( {\cal{R}}  )  )     \, + \, h.c. \, = \, 
e \, f(R) \, + \cdots\,\,,
\eea
where  $f(R)$ is given by
\bea
f(R) \, = \, \sum_m \, b_m \, R^{\, m+1}\,\,. 
\label{frgiv}
\eea
In this manner, given any $f(R)$ gravity specified by Eq (\ref{frgiv}),  we can supersymmetrize it by using the chiral action  (\ref{chiral} ), employing the function $ {\cal{F}} ({ \cal{R}},  T( {\cal{R}}  )  )  $ defined by  (\ref{kkk}).  According to   (\ref{hhh}) this is equivalent to a Lagrangian resulting by taking  $\omega = 0$ and 
\bea
 h ( \Phi, C )  &=&    {\cal{F}} ( -  \Phi / 2,  - 2 \, C   )  )   \,  = \, 
 6 \,  \sum_m \, b_m \,  {( -3 )}^{\,m}   \,  { (  - \Phi / 2 )  } \, { ( - 2 \, C ) }^{\, m}  \, = \, 
 - 3 \, \Phi \, \dfrac{ f(R) }{  R } \biggr\rvert_{R = 6 C}
\label{hhh2}
\eea
Therefore, the Lagrangian (\ref{Lagfin}) with
\bea
\omega \, = \, 0 \quad , \quad h(\Phi, C) \, = \,  - \Phi \, F(C)
\label{we1x}
\eea
or, due to the shift property, with 
\bea
\omega \, = \, F(C)+\bar{F}(\bar{C})         \quad , \quad h(\Phi, C) \, = 0\,\,,
\label{we2}
\eea
should embed an $f(R)$ gravity given by  
\bea
f(R) \, = \,  \dfrac{R}{3} \, F( {R}/{6} )\,\,.
\label{emb}
\eea
Note that    (\ref{we1x}) is exactly the minimal model studied before in (\ref{we11})  and, indeed, the $f(R)$  terms (\ref{emb}) are identical to the ones obtained in (\ref{frmin2}), as they should. This example is given in support of the statement that the chiral Lagrangians (\ref{chiral}) can  follow as limiting cases of the Lagrangians (\ref{Lagfin}) studied here. As a last remark,  the chiral Lagrangians are known to be plagued by ghosts.  One needs to depart from this description in order 
to possibly overcome the  problem of ghost states.  The framework of the theories given by (\ref{Lagfin}), being more general, offers an alternative approach to this issue, which might be fruitful.

\section{$N=1$ supergravities in the Einstein frame}

Working in the Einstein frame, in the usual description of the of the $N=1$ supergravity, 
the K\"{a}hler potential is, as usual, given by (\ref{KKK}) 
and the ordinary supergravity action is expressed in terms of the function $ \cal{G} $, defined by (\ref{kaler}),  and its derivatives. In general,  we  consider models  in which in addition to the fields  $ T, C, Q, \Phi $, extra fields may participate, which we collectively denote by $S$. We  first consider minimal models with only $ T, C, Q, \Phi $ present. 

\subsection{Minimal models}

We assume that no additional chiral multiplets are present and therefore $\Omega$ and $W$ are given by (\ref{cec1}), 
\bea
\Omega &=&   T + \bar{T}  +  \left( Q \, \bar{\Phi}+ \Phi \, \bar{Q} \right) + \omega   \left( C, \bar{C}, \Phi, \bar{\Phi}  \right) 
\nonumber \\
W &=&   T \, \Phi + Q \, C    + h  \left( C, \Phi \right)
\label{cec1x}
\eea

In a $T,\,C,\,\Phi,\,Q$ field basis, the analytic form of the K\"ahler matrix
  $  {\cal K}_{i \bar{j} }   $  is 
  \be
  {\cal K}_{i \bar{j} }   \, = \, \dfrac{ 3 }{ \Omega^2 }  \, 
 \left(
 \begin{array}{ c   ccc}
 \quad 1 \quad & \omega_{\bar{C}} & \quad Q +  \omega_{\bar{\Phi}} &  \Phi   \\
 \cdot & \omega_{{C}} \, \omega_{\bar{C}} - \Omega \,  \omega_{C \bar{C}} \; &  
 ( Q +  \omega_{\bar{\Phi}} ) \, \omega_{C} - \Omega \, \omega_{C \bar{\Phi}}  & \Phi \, \omega_C  \\
  \cdot  & \cdot &    \quad { | \bar{Q} + \omega_\Phi |}^2  - \Omega \, \omega_{\Phi \bar{\Phi}}   &  \Phi \, ( \bar{Q} + \omega_\Phi  )  - \Omega    \\
   \cdot & \cdot &  \cdot  &  { | \Phi |}^2  \\
      \cdot  &   \cdot  &  \cdot  &  \cdot 
 \end{array}
 \right)
 \label{matrmin}
 \ee
where, in order to save space the matrix elements below the diagonal are not explicitly shown, since by Hermiticity $  {\cal K}_{j \bar{i} }  =  {\cal K}_{i \bar{j} }^{\,*}  $. The determinant of the K\"ahler metric is given by 
\bea
\det \, {\cal K}_{i \bar{j} }  \, = \, \dfrac{81}{\Omega^{\, 5}} \,  \,  \omega_{C \bar{C}} \,.  
\label{determin}
\eea
In order to have a nonvanishing determinant we should demand that $ \omega_{C \bar{C}} \neq 0  $. 
Although the absence of negative norm states (ghosts) has as a necessary requirement that the determinant is positive, this by itself is not sufficient and does not guarantee absence of ghosts, since we might have an even number of ghost states. Therefore, an examination of all eigenvalues is necessary, this being a rather difficult task to accomplish, in general. Note that the positivity condition of the determinant
$  \det \, {\cal K}_{i \bar{j} } > 0 $ can be trivially satisfied, yielding that $  \omega_{C \bar{C}} \, < 0  \, $, since $\Omega < 0$. 
On the other hand the requirement $ \omega_{C \bar{C}} \neq 0  $ entails that in general we deal with a nonlocal $f(R)$ supergravity, as explained in the previous sections.  Therefore in these minimal models locality will be spoiled if we demand 
$  \det \, {\cal K}_{i \bar{j} } > 0 $ unless additional fields are introcuded.  This is discussed later on in this section. 

The class  of the  minimal models discussed here includes the models studied in previous sections [see (\ref{we11}) and (\ref{we3}  and (\ref{we44})], whose dual forms were explicitly given. However in all cases  studied there $\omega$ is independent of $C, \bar{C}$ and  therefore $  \det \, {\cal K}_{i \bar{j} } = 0 $ due to the fact that the second row in (\ref{matrmin})  has zero entries. 
Since, in this case, the determinant is vanishing it is not invertible. Therefore, in order to proceed further, one needs to separate the zero modes from the remaining states. That done,  it is found that  the relevant  $3 \times 3$ submatrix of the K\"ahler metric, after subtracting the zero mode,  has negative determinant which means that at least one ghost state exists. 
 
 Therefore, these models, as they stand  are ill-defined and a procedure or a recipe for removing the ghost states is required. The situation is reminiscent of the problem of instabilities encountered in the simple supergravity extension of the Starobinsky model where one of the chiral fields appearing in the K\"{a}hler potential $\, -3\ln(T+\overline{T}-|S|^2)$, namely $S$,  has to be stabilized at the origin, this being achieved by including an additional $(\overline{S}S)^4$ term \cite{EKN,KALI}. A possible way out of this problem, in the specific model, will be discussed in the following subsection.

\subsection{A stabilization procedure in the minimal model }

In order to investigate how the problem of unphysical degrees of freedom, encountered in the minimal model,  may be resolved we depart from the minimal choice $\omega = 0$ in the way prescribed below. In particular we  consider the case 
\bea
\Omega\, \, &=& \,T+\overline{T}+Q\overline{\Phi}+\overline{Q}\Phi\,+\,\lambda\,\overline{C}C \nonumber  \\
W\, &=& \,T\Phi+QC-\Phi\,F(C) 
\eea
This  generalizes the minimal model considered earlier. In fact now $\omega =  \lambda\,\overline{C} C $, i.e.  is  nonvanishing and 
$ h \,= \, -\Phi \,F(C) $ retains the same form. Since $ \bar{C} C$ terms appear in $\omega$  the dual supergravity is nonlocal.  Locality is expected to be established when $\lambda$ tends to zero in which case the model becomes exactly the minimal model.  Note that 
the $\lambda$-term is used to avoid zeros in the K\"{a}hler metric, at the cost of introducing nonlocal terms in the dual Lagrangian. The limit of vanishing $\lambda$ will be considered at the end. 
Recall that with the above definitions we have
\bea
K\,=\,-3\ln ( - \Omega/3)\,=\,3\ln(3)\,-3\ln\left( -T- \overline{T}- Q\overline{\Phi}-\overline{Q}\Phi\,-\,\lambda\,\overline{C}C\right)
\label{KAL}
\eea
At this point, following \cite{EKN}, we may introduce additional terms in $\Omega$ of the form $(\overline{\Phi}\Phi)^n$ with $n>2$. Such a term has the effect of {\textit{stabilizing}} $\Phi$ in the origin, i.e. at the value $\Phi=0$.
The K\"{a}hler metric $K_{i\bar{j}}$, corresponding to ({\ref{KAL}}) is, in a $\left({{T,\,C,\,\Phi, Q}}\right)$ basis
\bea
 \dfrac{3}{\Omega^2} \,  \left(
\begin{array}{cccc}
1 \,&\, {\lambda C} \,&\,Q\,&\, {\Phi}\\
\,&\,&\,&\,\\
{\lambda\overline{C}}\,&\,- {\lambda}{\Omega}+{\lambda^2|C|^2}\,& \,{\lambda\overline{C}Q}  &\, {\lambda \overline{C}\Phi}\, \\
\,&\,&\,&\,\\
{\overline{Q}}\,&\,{\lambda C\overline{Q}}\,&\,{|Q |^2}\,&\,- {\Omega}+{{\Phi}  \overline{Q}}\\
\,&\,&\,&\,\\
{\overline{\Phi}}\,&\,{\lambda C\overline{\Phi}}\,&\,- {\Omega}+{\overline{\Phi} {Q}}\,&\,{|\Phi|^2}
\end{array}
\right)\,\,\Longrightarrow\,\frac{3}{\Omega^2}
\left(\begin{array}{cccc}
1\,&\,\lambda C\,&\,Q\,&\,0 \\
\,&\,&\,&\,\\
{{\lambda\overline{C}}}\,&\,-\lambda(T+\overline{T})\,&\,\lambda\overline{C}Q  \, & 0 \\
\,&\,&\,&\,\\
\overline{Q} \,&\,\lambda {C} \overline{Q}   \,&\,|Q|^2 \,&\,-\Omega\\
\,&\,&\,&\,\\
0 \,&\,0 \,&\,-\Omega\,&\, 0
\end{array}\right)
\eea
where in the $\Phi\rightarrow 0$ limit $\Omega$ stands for  $\Omega\,=\,T+\overline{T}+\lambda|C|^2$.
The inverse matrix is
\bea
(K^{-1})^{\overline{i}j} \,=\, 
\,=\,\frac{\Omega}{3} \left(\begin{array}{cccc}
T+\overline{T}\,&\,{C}\,&\,0 \,&\, Q \,\\
\,&\,&\,&\,\\
\overline{C}\,&\,-\frac{1}{\lambda}\,&\,0\,&\,0\,\\
\,&\,&\,&\,\\
0 \,&\,0\,&\,0\,&\,-1\,\\
\,&\,&\,&\,\\
\overline{Q}  \,&\,0\,&\,-1\,&\,0\,
\end{array}\right)
\eea
The kinetic part of the Lagrangian {\textit{{in the linearized approximation}} is obtained by taking to zero nonlinear term, i.e. $C,\,Q\rightarrow 0$, except the $T$ which would create a zero eigenvalue. It is 
\be 
{\cal{L}}_{kin}\,\approx\,-\frac{3|\nabla T|^2}{(T+\overline{T})^2}\,+\frac{3\lambda |\nabla C|^2}{(T+\overline{T})}+\frac{3}{(T+\overline{T})}\left(\nabla \overline{Q}\cdot\nabla\Phi\,+\,\nabla Q\cdot\nabla\overline{\Phi}\right)\,.
\ee
It is clear that one of the combinations of $Q\pm\Phi$ is a ghost, since
$$\nabla Q\cdot\nabla\overline{\Phi}\,+\nabla\overline{Q}\cdot\nabla\Phi\,=\,\frac{1}{2}\left(|\nabla(Q+\Phi)|^2\,- |\nabla(Q-\Phi)|^2\right)$$
Nevertheless, removing $\Phi$  removes $Q$ as well, since the above mixed term is
$-3\Phi\nabla\cdot\left(\nabla \overline{{{Q}}}/(T+\overline{T})\right)+h.c.$.
The fact that the determinant equals  ${81 \lambda} \, {  {(T+\overline{T})}^{-5}}$ implies that there may be a second ghost present, namely $C$ in this approximation. This occurs  when $\lambda < 0$ due to the fact that $ T+\overline{T} $ should be negative 
($\Omega < 0$). At any rate, whatever the case,   this field is  removed in the $\lambda\rightarrow 0$ limit as we  see below.

The scalar potential
$$
V\,=\,e^{K}\,\left[\,G_{\overline{i}} \, (K^{-1})^{\overline{i} j} \, G_j\,-3|W|^2\right]\,\,{\text{with}}\,\,\,\,\,
G_i=W_i+WK_i,\,
$$
in the $\Phi=0$ limit, is 
\be
V\,=\,\frac{9}{(T+\overline{T}+\lambda|C|^2)^2} \left( \,\frac{|Q|^2}{\lambda}+\overline{T}C-\overline{F}(\overline{C})C+T\overline{C}-F(C)\overline{C}\,\right)\,.
\ee
On the other hand, the kinetic Lagrangian is
$$
{\cal{L}}_{kin}\,=\,-\frac{3|\nabla T|^2}{(T+\overline{T})^2}\,+\frac{3\lambda |\nabla C|^2}{(T+\overline{T})}
\,.$$
Since, no term for $Q$ is present in the kinetic part, its equation of motion is just
$$\frac{\partial V}{\partial Q}\,=\,Q^*\,\frac{\partial V}{\partial |Q|^2}\,=\,0\,,$$
which is satisfied with $Q=0$. Therefore, taking $Q=0$, we arrive at a theory of $T$ and $C$ only. Now we may take the limit $\lambda=0$. In this limit $C$ becomes auxiliary. The pertinent Lagrangian is 
\bea
{\cal{L}} \, = \, 
-\frac{3|\nabla T|^2}{(T+\overline{T})^2}\,-\frac{9}{(T+\overline{T})^2}\left(T\overline{C}+\overline{T}C-C\overline{F}(\overline{C})-\overline{C}F(C)\right)\,.
\eea
Since, $C$ has no kinetic term, we obtain $C$ as the solution of
\be
\overline{C}\frac{dF(C)}{dC}\,+\, {\overline{F}}(\overline{C})\,=\,\overline{T}
\,.
\ee
For the sake of simplicity we may ignore the imaginary parts.
\footnote{
Including the imaginary parts and defining $T\,=\,t\,+\,i\alpha$ and $C\,=\,c\,e^{i\beta}$ and $F(C)\,=\,\sum_{n=0}f_nc^ne^{in\beta}$,
the Lagrangian takes on the form
$${\cal{L}}\,=\,-\frac{3}{4t^2}\left((\nabla t)^2+(\nabla\alpha)^2\right)\,-\frac{9}{2t^2}\left(ct\cos\beta+\alpha c\sin\beta\,-c\sum_{n=0}f_nc^n\cos((n-1)\beta)\,\right)\,.$$
Minimizing with respect to $\alpha$ and $\beta$, we obtain
$$
\frac{\partial V}{\partial\alpha}=0\,\Longrightarrow\,c\sin\beta\,=\,0\,,\,\,\,\,\,\,\,\,\,
\frac{\partial V}{\partial\beta}\,=\,0\,\Longrightarrow\,\,-ct\sin\beta+\alpha c\cos\beta+c\sum_{n=0}f_nc^n(n-1)\sin((n-1)\beta)\,=\,0$$
which can be satisfied for $\alpha=\beta=0\,.$} 
Then, we have
\be
{\cal{L}}\,=\,-\frac{3}{4}\frac{(\nabla T)^2}{T^2}\,-\frac{9}{2T^2}\left(CT-CF(C)\right)
\,.
\ee
Minimizing with respect to $C$ we obtain
$$T=\frac{d}{dC}(CF(C))\,=\,{{\left(CF(C)\right)'}}$$
We may introduce
\be 
f(C)\,=\,CF(C)\,.
\ee
Then, we have
\bea
{\cal{L}}\,=\,-\frac{R}{2}\,-\frac{3}{4}\frac{(\nabla f')^2}{(f')^2}\,-\frac{9}{2}\frac{(Cf'-f)}{(f')^2}
\label{eeiinn}
\eea
Compare this to a Jordan-theory
\bea  
   {\cal{L}}_J   \, = \, 
-\frac{\sigma f'(C)}{2}R\,-\frac{9 \, \sigma^2}{2}(Cf'(C)-f(C) )
\label{jjjo}
\eea
which in the Einstein frame becomes identical to our Einstein theory (\ref{eeiinn}).  Thus our theory   corresponds to the class 
of Lagrangians (\ref{jjjo}), related by constant Weyl rescalings,  with the constant  $\sigma$ undetermined. 
On the other hand, if we eliminate $C$ from the Jordan-theory we obtain
$$
\frac{\delta {\cal{L}}_J}{\delta C}\,=\,0\,\Longrightarrow\,f''(C)\left(C+\frac{R}{9 \, \sigma } \right)\,=\,0\,\,\Longrightarrow\,{{{\cal{L}}_J\,=\,
\left.\frac{9 \, \sigma^2}{2} \, f(C)\right|_{C = - {R} / \,  {9 \, \sigma}}}}
$$
This is a genuine ${\cal{F}}(R)$ theory defined as
\bea
{ {{\cal{L}}_J\,=\,{\cal{F}}(R)\,\equiv\,\frac{9 \, \sigma^2}{2} \, f(- R / \, 9 \,  \sigma)\,=\,-\frac{\sigma \, R}{2} \, F(-R / \, 9 \, \sigma)\,.}}
\eea
Choosing $\sigma=-2/3$ we arrive at the results  obtained in ({\ref{frmin2}}) and ({\ref{emb}}), 
$$
-\frac{\sigma \, R}{2} \, F(-R / \, 9 \, \sigma) \, = \, \frac{ \, R}{3} \, F( R / \, 6 \, )
$$

Note that the choice $F(C)\,=\,AC+B$ leads, through minimization, to
$$
V(T)\,=\,\frac{9}{2A}\frac{|T-B|^2}{(T+\overline{T})^2}
$$
which coincides with the Supergravity generalizations of the Starobinsky model encountered in previous works \cite{KTAL}. Indeed the dual form for $F(C)=AC+B$  is
$${\cal{L}}\,=\,\frac{3B}{2}C\,-\frac{3A}{2}C^2\,=\,-\frac{B}{2}R\,+ \frac{A}{18} \, R^2$$
which requires $B=1$ and $A > 0$. Another example is given by the choice $F(C)\,=\,A\,+\,B\,C^N$, which leads to
\bea
{\cal{L}}\,= \,- \dfrac{A}{2} \,R\,- \, \dfrac{B}{2} \,   \dfrac{ {  ( - 1 )  }^N  }{  9^N }    \,R^{N+1} \, \sim \, {\cal{L}}\,\sim\,-R\,+\,R^{N+1}\,.
\eea

\subsection{Nonminimal models with extra fields}

The minimal models discussed before have flaws that may be possibly circumvented if additional fields are present. These models are more general and in particular limits, where the extra degrees of freedom are frozen, they  may yield the minima cases discussed before. 
Therefore, we  consider models  in which apart  the fields  $ T, C, Q, \Phi $, additional fields  participate, denoted collectively  by $S$. For definiteness we  assume the existence of only one field $S$ and the generalization to an arbitrary number of them can be done in a straightforward manner. 
Labeling again the fields as $ 1 = T, 2 = C, 3 = \Phi, 4 = Q $ and $5 = S$, the analytic form of the matrix
  $  {\cal K}_{i \bar{j} }   $  is 
  \bea
  &&  {\cal K}_{i \bar{j} }   \, = \, \dfrac{ 3 }{ \Omega^2 }  \, \times  \nonumber \\
  && \left(
 \begin{array}{ ccccc}
 \quad 1 \quad & \omega_{\bar{C}} & \quad Q +  \omega_{\bar{\Phi}} &  \Phi  &  \omega_{\bar{S}}  \\
 \cdot & \omega_{{C}} \, \omega_{\bar{C}} - \Omega \,  \omega_{C \bar{C}} \; &  ( Q +  \omega_{\bar{\Phi}} ) \, \omega_{C} - \Omega \, \omega_{C \bar{\Phi}}  & \Phi \, \omega_C & \omega_{{C}} \, \omega_{\bar{S}} - \Omega \,  \omega_{C \bar{S}}  \\
  \cdot  & \cdot &    \quad { | \bar{Q} + \omega_\Phi |}^2  - \Omega \, \omega_{\Phi \bar{\Phi}}   &  \Phi \, ( \bar{Q} + \omega_\Phi  )  - \Omega   &  \; \bar{Q} \, \omega_{\bar{S}}  +   \omega_{{\Phi}} \, \omega_{\bar{S}} - \Omega \,  \omega_{\Phi \bar{S}} \\
   \cdot & \cdot &  \cdot  &  { | \Phi |}^2 & \bar{\Phi} \, \omega_{\bar{S}} \\
      \cdot  &   \cdot  &  \cdot  &  \cdot  &   \omega_{{S}} \, \omega_{\bar{S}} - \Omega \,  \omega_{S \bar{S}}
 \end{array}
 \right)
 \label{matr}
 \eea
Again, the matrix elements below the diagonal are not explicitly shown, since they are the complex conjugates of the elements lying above the diagonal,  $  {\cal K}_{j \bar{i} }  =  {\cal K}_{i \bar{j} }^{\,*}  $, the matrix ${\cal K}_{j \bar{i} }    $ being Hermitian. The determinant of the K\"ahler metric in this case is given by 
\bea
\det \, {\cal K}_{i \bar{j} }  \, = \, \dfrac{243}{\Omega^{\, 6}} \, ( \,  \omega_{S \bar{C}}  \,  \omega_{C \bar{S}} \, -  \omega_{C \bar{C}} \,  \omega_{S \bar{S}}  \, )\,\,. 
\label{deter1}
\eea
To avoid ghosts it is mandatory that  the determinant  is positive. However, as already discussed in the minimal case,  this does not ensure absence of ghost states, since the presence of an even number of ghosts cannot be excluded.  Therefore a thorough examination of all eigenvalues seems necessary. 

In the following section we  consider special models in which the additional fields $S$ participate in a particular manner.

\subsection{Sequestered models}

The treatment of additional fields $S$ is facilitated if they participate in a sequestered manner. This means that their involvement is implemented in a way that allows them not to  mix with $T, C, \Phi, Q$ within the function $\Omega$ and the superpotential $W$.  This class of models is characterized by the functions defined in  (\ref{cec1}) being of the following form 
\be
\begin{array}{l}
\omega   \left( C, \bar{C}, \Phi, \bar{\Phi} , S, \bar{S} \right) \, =\,  N( C, \bar{C}, \Phi, \bar{\Phi}   ) + \Sigma( S, \bar{S} )  \\
\,\\
 h  \left( C, \Phi, S \right)  \, = \, f(C,\Phi ) + P( S) 
\end{array}
\label{cec2}
\ee
i.e. the role of fields $S$ is sequestered from that of the remaining fields within the function $\Omega$ and the superpotential.  

The dual description of this model can be read from (\ref{Lagfin}), which due to the sequestered character of the model can be shown to lead to a Lagrangian of the following form 
\bea
 e^{-1} \, {\cal{L}}_0 \, = \,   e^{-1} \, {\cal{L}}_{\Phi,C} +  e^{-1} \, {\cal{L}}_S\,,
 \label{eqlag}
\eea
where ${\cal{L}}_{\Phi,C } $  depends on  the fields $\Phi, C$ but not $S$, while the second term ${\cal{L}}_S $ includes the dependence on $S$, as well as other fields. The analytic form of ${\cal{L}}_{\Phi,C}$ is given by 
\bea
e^{-1} \, {\cal{L}}_{\Phi, C} \, &=& \,   
   \dfrac{N }{6}  \,  \left( \  R +  \frac{2}{3} M \bar{M} - \frac{2}{3} b_{\mu} b^{\mu}\right) 
-  N_{a \bar b}  \, \partial_{\mu} \phi^a \partial^{\mu}{\bar \phi}^{\bar b} + N_{a \bar b} \, F^a F^{\bar b} - 
\frac{i}{3} \left( N_a \, \partial_{\mu} \phi^{a} 
 - N_{\bar a} \, \partial_{\mu} {\bar \phi}^{\bar a} \right) b^{\mu}  \nonumber \\
&& -  \frac{M}{3}  \, \left( N_a F^a + 3 \bar f \right)  -  \frac{\bar M}{3} \left( \omega_{\bar a} {\overline F}^{\bar a} + 3 f \right) + 
f_a F^a + {\bar f} _{\bar a}  {\overline F}^{\bar a}\,.
\label{Lagfin2}
\eea
In this, the indices $a,b$ run over $c, \phi$ and  $M, c, F^c, F^\phi$ are given by (\ref{25a} - \ref{25d}).  The $S$-dependent part is 
given by
\bea
e^{-1} \, {\cal{L}}_S \, &=& \,   
   \dfrac{\Sigma }{6}  \,  \left( \  R +  \frac{2}{3} M \bar{M} - \frac{2}{3} b_{\mu} b^{\mu}\right) 
-  \Sigma_{s \bar s}  \, \partial_{\mu} s \, \partial^{\mu}{\bar s} + \Sigma_{s \bar s} \, F^s F^{\bar s} - 
\frac{i}{3} \left( \Sigma_s \, \partial_{\mu} s
 - { \Sigma }_{\bar s} \, \partial_{\mu} {\bar s} \right) b^{\mu}  \nonumber \\
&& -  \frac{M}{3}  \, \left( \Sigma_s \,  F^s + 3 \bar P \right)  -  \frac{\bar M}{3} \left( \Sigma_{\bar s} {\overline F}^{\bar s} + 3 P \right) + 
P_s \, F^s + {\bar P} _{\bar s} \, {\overline F}^{\bar s}\,.
\label{Lagfin3}
\eea
In this, $F^s$ is given by [see Eq. (\ref{fss})] 
\bea
{\overline{F}}^{\bar{s}} \, = \,  \Sigma_{s \bar{s}}^{-1} \, \left(  \phi \, \Sigma _s - P_s  \, \right)\,,
\label{fss2}
\eea
so that upon plugging  (\ref{fss2}) into (\ref{Lagfin3}) we get 
\bea
e^{-1} \, {\cal{L}}_S \, &=& \,   
   \dfrac{\Sigma }{6}  \,  \left( \  R +  \frac{2}{3} M \bar{M} - \frac{2}{3} b_{\mu} b^{\mu}\right) 
-  \Sigma_{s \bar s}  \, \partial_{\mu} s \, \partial^{\mu}{\bar s} - 
\frac{i}{3} \left( \Sigma_s \, \partial_{\mu} s
 - { \Sigma }_{\bar s} \, \partial_{\mu} {\bar s} \right) b^{\mu}  \nonumber \\
&& -   \, 3 \, {( \,  \phi \, \overline{P} +{\overline  \phi} \, P  \, )} \, 
- \, \Sigma_{s \overline{s}}^{-1} \,  {|  \,  \phi \, \Sigma_s - P_s \,  |}^{\,2} \, 
\label{Lagfin3xx}
\eea
The Lagrangian (\ref{eqlag}), with ${\cal{L}}_{\Phi,C} \, , \,  {\cal{L}}_{S} $ defined by (\ref{Lagfin2}) and (\ref{Lagfin3xx}) respectively, is the  higher derivative  dual form of the ordinary $N = 1 $ supergravity theory corresponding to the choice 
(\ref{cec2}).

From the point of view of the ordinary supergravity, the 
 K\"ahler metric (\ref{matr}) takes a rather  simpler form given by 
  \bea
  &&  {\cal K}_{i \bar{j} }   \, = \, \dfrac{ 3 }{ \Omega^2 }  \, \times  \nonumber \\
  && \left(
 \begin{array}{ c   cccc}
 \quad 1 \quad & N_{\bar{C}} & \quad Q +  N_{\bar{\Phi}} &  \Phi  &  \Sigma_{\bar{S}}  \\
 \cdot & N_{{C}} \, N_{\bar{C}} - \Omega \,  N_{C \bar{C}} \; &  ( Q +  N_{\bar{\Phi}} ) \, N_{C} - \Omega \, N_{C \bar{\Phi}}  & \Phi \, N_C & N_{{C}} \, \Sigma_{\bar{S}}  \\
  \cdot  & \cdot &    \quad { | \bar{Q} + N_\Phi |}^2  - \Omega \, N_{\Phi \bar{\Phi}}   &  \Phi \, ( \bar{Q} + N_\Phi  )  - \Omega   &  
  \;(  \bar{Q} \,  +   N_{{\Phi}} \, ) \Sigma_{\bar{S}} \\
   \cdot & \cdot &  \cdot  &  { | \Phi |}^2 & \bar{\Phi} \, \Sigma_{\bar{S}} \\
      \cdot  &   \cdot  &  \cdot  &  \cdot  &   \Sigma_{{S}} \, \Sigma_{\bar{S}} - \Omega \,  \Sigma_{S \bar{S}}
 \end{array}
 \right)
 \label{matr2}
 \eea
In this case the determinant of the K\"ahler metric, obtained from ( \ref{deter1} ),  is just
\bea
det \, {\cal K}_{j \bar{i} }  \, = \, - \, \dfrac{243}{\Omega^{\, 6}} \,    N_{C \bar{C}} \,  \Sigma_{S \bar{S}}\,.  
\label{deter2}
\eea
Therefore, we need $  N_{C \bar{C}} \,  \Sigma_{S \bar{S}}  \neq 0 $ for nonvanishing determinant, while we must have
$  N_{C \bar{C}} \,  \Sigma_{S \bar{S}}  < 0 $ for it to be positive. Nevertheless, as we repeatedly mentioned this does not exclude the possibility of having an even number of ghost states. 

A complete study of the mass  spectrum in an arbitrary background is difficult to carry out. In order to proceed further we  explore the spectrum assuming an expansion of the scalar kinetic terms 
 around specific background solutions for which $\Sigma_S=0$ but $\Sigma_{S \bar{S}} \neq 0  $. Then,  the field $S$ does not mix with the rest of the fields in the kinetic terms, and only the $4 \times 4$ submatrix of   $  {\cal K}_{i \bar{j}}  $ matters.  
If,  moreover, we assume that the backgrounds under discussion are such that  $ Q = \Phi = 0$,  and the functions $N, \Sigma$ on these backgrounds are such that   $N_\Phi = N_C = N_{C \bar{\Phi}} = 0  $,  then, the kinetic terms simplify a great deal. In particular, in this case  the K\"ahler metric is given by 
  \bea
  &&  {\cal K}_{i \bar{j} }   \, = \, \dfrac{ 3 }{ \Omega^2 }  \, \times  \nonumber \\
  && \left(
 \begin{array}{ c   cccc}
 \quad 1 \quad & 0 &  0  &  0  &  0  \\
0 &  - \Omega \,  N_{C \bar{C}} \; &  0  & 0  & 0  \\
 0  & 0 &     - \Omega \, N_{\Phi \bar{\Phi}}   & - \, \Omega  &   0 \\
  0 & 0 &  - \, \Omega  &  0  & 0  \\
     0  &   0  &  0 & 0  &    - \, \Omega \,  \Sigma_{S \bar{S}}
 \end{array}
 \right)
 \label{matr3}
 \eea
so that  the kinetic terms are
\be
-\dfrac{ 3 }{ \Omega^{\, 2} }  \,\bigg(   
\partial_\mu T  \partial^\mu \bar{T} - \Omega   N_{C \bar{C}}  \partial_\mu C   \partial^\mu \bar{C}    -   
\Omega   \Sigma_{S \bar{S}}  \partial_\mu S   \partial^\mu \bar{S} 
 -  \Omega  N_{\Phi \bar{\Phi}}    \partial_\mu \Phi   \partial^\mu \bar{\Phi}  -  
   \Omega   \partial_\mu \Phi   \partial^\mu \bar{Q}  -  
    \Omega   \partial_\mu Q   \partial^\mu \bar{\Phi} 
\bigg)\,.
\label{kinet}
\ee
In this, all quantities are meant at the background values and, therefore $  N_{\Phi \bar{\Phi}} , N_{C \bar{C}} ,  \Sigma_{S \bar{S}} \, $ are essentially constants. Note that  
this is not completely diagonal since there is a mixing in the $Q, \Phi$ sector.  One, and only one,  of the eigenstates of this sector is a ghost and this  is independent of the sign of the product $ N_{C \bar{C}} \,  \Sigma_{S \bar{S}} $. 
Thus, there is at least one ghost state. This is the unique ghost state  provided $ N_{C \bar{C}} > 0 , \Sigma_{S \bar{S}} > 0  $. Recall that $\Omega < 0$.  However, if   $ N_{C \bar{C}} <  0 , \Sigma_{S \bar{S}} < 0  $, the states $C,S$ are ghosts too and we have three ghosts. In any other case we have two ghost states, one from the  $Q, \Phi$ sector and another is either $C$ or $S$.  We pursue  the case of having just one ghost state. To this purpose we diagonalize the $Q , \Phi  $ kinetic  submatrix,  defining the fields $G$ and $H$, which are related to $\Phi, Q$ by the following unitary transformation
\be
G \, = \,\dfrac{\Phi - \lambda_+ \, Q}{\sqrt{ 1 \, + \,  \lambda_+^2 }}\,,\,\,\,\, \,
H \,= \,  \dfrac{\Phi - \lambda_- \, Q}{\sqrt{  1 \, + \,  \lambda_-^2} }
\label{ghost}\,.
\ee
The quantities $\lambda_{\pm}$ are proportional to the eigenvalues of the $\Phi, Q$ submatrix and are given by
\bea
\lambda_{\pm} \, = \, \frac{1}{2}\left({\cal{N}} \pm \sqrt{ {\cal{N}}^2 + 4 } \right)\,, 
\eea
where $ {\cal{N}} \equiv   N_{\Phi \bar{\Phi}}  $, treating its value at the background solution as a constant. The eigenvalue $\lambda_{-}$ is always  negative.  
In this manner, the kinetic terms given in Eq. ( \ref{kinet} ) are given by 
\be
- \, \dfrac{ 3 }{ \Omega^{\, 2} }  \,\bigg(    \, 
\partial_\mu T \, \partial^\mu \bar{T} - \Omega \,  N_{C \bar{C}} \, \partial_\mu C  \, \partial^\mu \bar{C} \,   - \,  
\Omega \,  \Sigma_{S \bar{S}} \, \partial_\mu S  \, \partial^\mu \bar{S}  
 - \, \Omega \, \lambda_+ \,  \partial_\mu H  \, \partial^\mu \bar{H} \, 
- \, \Omega \, \lambda_- \,  \partial_\mu G  \, \partial^\mu \bar{G} \,  \quad 
\bigg)
\label{kinet2}
\ee
$G$ is the ghost field in this case, since $\lambda_-  $ is negative.  Along similar lines as above, arguments have been invoked in support of  the statement that in this class of models the existence of at least one ghost state in unavoidable.  However, it may happen that  this state decouples from the remaining physical states in which case  the model is viable and should not be abandoned \cite{CECOTTI}. 

In order to be more specific let us explore a model in which
\be
N( C, \bar{C}, \Phi, \bar{\Phi}   ) \, = \, \mu \, {| C |}^{\,2} \, + \, \lambda \,    {| C |}^{\, 4} \,,\,\,\,\,\,\,
 \Sigma( S, \bar{S} ) \, = \, \lambda_1 \, {| S |}^{\,2} \, + \, \lambda_2 \,    {| S |}^{\, 4}\,\,. 
 \label{spec}
\ee
The superpotential has the general form  given by (\ref{cec2}). In this case due to (\ref{deter2}) we need both $\mu$ and $\lambda_1$ to be nonvanishing in order to have a nonvanishing K\"ahler metric.  Along the direction $S = Q = C = \Phi = 0  $ and taking 
$ Im \, T =0  $, the bosonic part of the Lagrangian becomes
\be
- \, \frac{3}{4}  \frac{( \nabla  Re \, T )^2  }{ (Re \, T)^2 } \, - \, 
\frac{9}{4 (Re \, T)^2  } \, \left( \,     \dfrac{ { | P^\prime(0) |}^{\, 2} }{\lambda_1  }  \, +  \dfrac{1}{\mu} \,   
{ \Biggr\rvert \dfrac{ \partial f (0,0) }{\partial \, C } \Biggr\rvert  }^2   \,  \right)\,. 
\ee
Using the canonically normalized field $\varphi$ 
\bea
Re \, T \, = \, - \, e^{\sqrt{\frac{2}{3} } \, \varphi}
\label{normf}
\eea
the above Lagrangian becomes 
\bea
- \, \dfrac{1 }{2}  \, { \partial_\mu \varphi  } \, { \partial^\mu \varphi } \, - \, 
\dfrac{9}{4 \,  } \, \left( \,     \dfrac{ { | P^\prime(0) |}^{\, 2} }{\lambda_1  }  \, +  \dfrac{1}{\mu} \,   
{ \Biggr\rvert \dfrac{ \partial f (0,0) }{\partial \, C } \Biggr\rvert  }^2   \,  \right) \,
\exp{ \left[  - 2 \, \sqrt{\frac{2}{3} } \, \varphi \right]} 
\eea
The minus sign in (\ref{normf}) reflects the fact that $\Omega $ must be negative, therefore  when all other fields are set to zero the real part of $T$ must be negative too. 
Obviously the scalar potential above  is not of the Starobinsky type and deformations of the superpotential are needed in order to get Starobinsky-like potentials suitable for inflation. Nevertheless, such  considerations lie beyond the scope of the present  work. 


\section{Summary and Conclusions}

The phenomenological success of the Starobinsky model of inflation offers ample motivation to consider its embedding in more general schemes like a supergravity theory, as well as considering its generalization to a gravity theory in which the scalar curvature does not appear only linearly in the action but through a more general function $f(R)$. The latter type of theories can easily be shown to be equivalent to Einstein gravity coupled to a scalar field whose interactions are determined by the function $f$. The task of promoting this scheme to a supersymmetric construction and, thus, formulating an $f(R)$ supergravity theory is not straightforward. This is exactly the task we undertook in this paper and tried to carry  a step further. We started with the Jordan-frame supergravity coupled to a number of, at least four, chiral superfields. Two of these fields appear linearly in the specifically chosen kinetic function
 $\Omega = T+\overline{T}+ Q \overline{\Phi}+\overline{Q} \Phi+\omega(C,\overline{C},\Phi,\overline{\Phi}, \cdots)$ and the superpotential, $W=T \Phi+Q C + h(C,\Phi, \cdots)$.  
 Nevertheless, this construction is general enough to include classes of models, parametrized through the two functions $\omega,\,h$. In the resulting theory gravity enters through a general function of the curvature. Depending on the chosen $\omega$ term in the kinetic function, this can contain derivatives of the curvature or not. Thus, in the local case, the bosonic part of this supergravity theory is a gravity theory of the $f(R)$ type. We analyzed specific classes of models, corresponding to choices of $\omega$ and $h$. We also showed that chiral Lagrangian models correspond to special cases in our general framework. Next, we studied the corresponding Einstein-frame theory formulated in the standard way in terms of a number of chiral superfields coupled through a K\"{a}hler potential and a superpotential chosen as above. We discussed the issue of ghost states and, although, no general resolution was offered, we gave possible ways that unphysical states could decouple within specific models. Summarizing, we have presented a particular framework for a $f(R)$ supergravity theory and analyzed specific models. Despite still open issues like the absence of unphysical states the framework is general enough to render its applicability fruitful in the construction of specific models of inflation beyond the Starobinsky model.

\vspace*{5mm}  
{\textbf{ACKNOWLEDGMENTS}}   

This research has been cofinanced by the European Union (European Social Fund - ESF)
and Greek national funds through the Operational Program Education and Lifelong
Learning of the National Strategic Reference Framework (NSRF) - Research Funding
Program: {\textit{ARISTEIA-Investing in the society of knowledge through the European Social Fund.}} One of us (K.T.) would also like to thank I. Antoniadis and K. Papadodimas for discussions and hospitality at the CERN Theory Division.

\end{document}